\documentclass[twocolumn,prl,aps,showpacs,superscriptaddress]{revtex4-1}
\usepackage{graphicx,graphics}
\usepackage{amsmath}
\usepackage{amssymb}
\usepackage{epstopdf}
\usepackage{amstext}
\usepackage{amsthm}
\usepackage{amsfonts}
\usepackage{latexsym}
\usepackage{array}
\usepackage{xfrac}
\usepackage{lipsum}
\usepackage[toc,page]{appendix}
\usepackage{subfigure}
\newcommand{\igbjd}[1]{}\newcommand{\beqa}{\begin{eqnarray}}
\newcommand{\eeqa}{\end{eqnarray}}
\newcommand{\beq}{\begin{equation}}
\newcommand{\eeq}{\end{equation}}
\usepackage[normalem]{ulem}

\usepackage[colorlinks=true,citecolor=Cerulean,linkcolor=RubineRed,urlcolor=Cerulean,pdftex]{hyperref}

\usepackage{xcolor}
\definecolor{Cerulean}{rgb}{0.,0.59,0.835}
\definecolor{RubineRed}{rgb}{0.61,0.07,0.12}

\begin{document}

\title{Universal dimerized quantum droplets in a one-dimensional lattice}
\author{Ivan Morera}
\affiliation{Departament de F\'isica Qu\`antica i Astrof\'isica, 
Facultat de F\'{\i}sica, Universitat de Barcelona, E--08028 Barcelona, Spain}
\affiliation{Institut de Ci\`encies del Cosmos, Universitat de Barcelona, 
ICCUB, Mart\'i i Franqu\`es 1, Barcelona 08028, Spain}
\author{Grigori E. Astrakharchik}
\affiliation{Departament de F\'isica, Universitat Polit\`ecnica de Catalunya, 
Campus Nord B4-B5, E-08034 Barcelona, Spain}
\author{Artur Polls}
\affiliation{Departament de F\'isica Qu\`antica i Astrof\'isica, 
Facultat de F\'{\i}sica, Universitat de Barcelona, E--08028 Barcelona, Spain}
\affiliation{Institut de Ci\`encies del Cosmos, Universitat de Barcelona, 
ICCUB, Mart\'i i Franqu\`es 1, Barcelona 08028, Spain}
\author{Bruno Juli\'{a}-D\'{i}az}
\affiliation{Departament de F\'isica Qu\`antica i Astrof\'isica, 
Facultat de F\'{\i}sica, Universitat de Barcelona, E--08028 Barcelona, Spain}
\affiliation{Institut de Ci\`encies del Cosmos, Universitat de Barcelona, 
ICCUB, Mart\'i i Franqu\`es 1, Barcelona 08028, Spain}

\date{\today}

\begin{abstract}
The ground-state properties of two-component bosonic mixtures in a 
one-dimensional optical lattice are studied both from few- and 
many-body perspectives. We rely directly on a microscopic Hamiltonian 
with attractive inter-component and repulsive intra-component 
interactions to demonstrate the formation of a quantum liquid. We 
reveal that its formation and stability can be interpreted in terms 
of finite-range interactions between dimers. We derive an effective 
model of composite bosons (dimers) which correctly captures both the 
few- and many-body properties and validate it against exact results 
obtained by DMRG method for the full Hamiltonian. The threshold for 
the formation of the liquid coincides with the appearance of a bound 
state in the dimer-dimer problem and possesses a universality in terms 
of the two-body parameters of the dimer-dimer interaction, namely 
scattering length and effective range. For sufficiently strong effective 
dimer-dimer repulsion we observe fermionization of the dimers which 
form an effective Tonks-Girardeau state. Finally, we identify conditions 
for the formation of a solitonic solution. 
\end{abstract}
\maketitle

{\bf Introduction.} 
A microscopic theory of liquids relies on the specific properties of 
the atom-atom interaction potential. In classical liquids the typical 
interaction potential has a van der Waals shape for which the long-range 
attraction is compensated by a short-range repulsion~\cite{hansen}. In 
fermionic quantum fluids, e.g. electron gas, neutron stars, etc, the 
Pauli exclusion principle naturally provides a hard-core short-range 
repulsion~\cite{fetter2003quantum}. The classical picture carries over to 
the bosonic quantum realm where microscopic descriptions of very 
different systems, e.g. liquid helium, require strong short-range 
repulsive forces~\cite{leggett2006quantum}. This paradigm has been recently 
challenged by the experimental 
observation~\cite{PhysRevLett.116.215301,Schmitt2016,PhysRevX.6.041039,
Cabrera301,PhysRevLett.120.135301,PhysRevLett.120.235301,PhysRevResearch.1.033155} 
of quantum droplets which stability is due to a compensation between the 
mean-field interactions and quantum fluctuations~\cite{PhysRevLett.115.155302,
PhysRevA.94.021602,PhysRevA.94.043618,Kartashov2019}. 

The dimensionality of the system has strong implications for the properties 
of these quantum droplets~\cite{PhysRevLett.117.100401}. In 
the one-dimensional (1D) case, droplets get formed in the regime where at the mean-field level 
the system is on average repulsive. That is, the quantum fluctuations result in 
an effective attraction which is able to liquefy the system~\cite{PhysRevLett.117.100401}. 
The properties of such one-dimensional liquids have been studied in the 
continuum~\cite{PhysRevLett.122.105302,ParisiGiorgini2020,Ota2020} and, recently, 
extended to optical lattices~\cite{PhysRevResearch.2.022008}. The latter are 
particularly appealing as the phenomena takes place at small filling fractions 
$\simeq 2$ which greatly increases the life-time of these droplets. Moreover, in 
the 1D case the low-density regime corresponds to stronger correlations which 
makes the few-body problem very interesting. 

In this Letter we concentrate on the discrete setup and describe the transition 
between the gas, liquid and soliton phases in a strongly interacting bosonic 
mixture in a one-dimensional optical lattice. Starting from a two-component 
Bose-Hubbard Hamiltonian, describing a bosonic mixture loaded in a  1D optical 
lattice, we derive an effective dimer model. We rely on it to write down 
explicit analytical expressions for the dimer-dimer scattering length and 
effective range. This allows us to analytically predict the tetramer bound 
state threshold and its binding energy. The obtained simple expressions are 
tested in a comparison with full density matrix renormalization group (DMRG) 
calculations of the original bosonic mixture and an excellent agreement is 
found. The gas to liquid transition of the original model is found to take 
place at the threshold for the formation of a bound tetramer, i.e. when 
the effective dimer-dimer interactions switch from repulsion (gas) to 
attraction (liquid). The effective dimer model is also able to explain the 
properties of the liquid phase observed in the many-body problem. We 
recognize that the stability of the liquid stems from the effective range 
contribution to the dimer-dimer scattering problem, contrarily to the 
scenario reported in continuum where the stabilization mechanism 
was instead 
attributed to three-dimer interactions~\cite{PhysRevA.97.063616,PhysRevA.97.061605}. 
Finally, we identify conditions necessary for soliton formation in this discrete system.

{\bf Model system.}
We study a binary mixture of bosons interacting via short-range interactions 
and loaded into a high 1D optical lattice at zero temperature. The system is 
described by the Bose-Hubbard Hamiltonian~\cite{lewenstein2012ultracold}
\begin{eqnarray}
H&=&-t \, \sum_i \sum_{\alpha=A,B} \left( \hat{b}_{i,\alpha}^{\dagger}\hat{b}_{i+1,\alpha}+\text{h.c.}\right) 
\label{eq:TwoBH}\\
&+&\frac{U}{2}\sum_i \sum_{\alpha=A,B} \left(\hat{n}_{i,\alpha}\left(\hat{n}_{i,\alpha}-1\right)\right)
+U_{ab}\sum_{i}\hat{n}_{iA}\hat{n}_{iB} \,,\nonumber
\end{eqnarray}
where $\hat{b}_{i\alpha}$ ($\hat{b}_{i\alpha}^{\dagger}$) are the annihilation (creation)  bosonic operators at site $i=1,\dots,L$ for species $\alpha=A,B$, respectively, and $\hat{n}_{i\alpha}$ are their corresponding number operators. The DMRG calculations will be performed using open boundary conditions and the typical number of sites will be $L=32$.
We consider a symmetric mixture with equal tunneling strength, $t>0$, repulsive intra-species interaction strength, $U>0$, equal for both components, and attractive inter-species interaction $U_{ab}<0$.

{\bf Effective dimer-dimer interaction.}
We start by addressing a few-body problem and consider four bosons $N_a=N_b=2$ described by the Hamiltonian~\eqref{eq:TwoBH}. 
A prominent feature of 1D geometry is that it facilitates the formation of $a$-$b$ dimers for inter-species attraction. In the regime of strong interactions, $|U_{ab}|/t \approx U/t \gg 1$ with $r\equiv (U+U_{ab})/U\ll 1$, each dimer gets localized on a single site, thus allowing the reduction of the Hilbert space to the dimer subspace~\cite{PhysRevLett.92.050402,PhysRevLett.92.030403,PhysRevLett.103.035304}. A similar approach has been employed to study the trimer problem in the single-component Bose-Hubbard model~\cite{PhysRevA.81.011601}. 
As well, 
the initial four-body problem can be reduced to a two-body problem of dimers described by an effective Hamiltonian (refer to Supplemental material for a detailed derivation), 
\begin{eqnarray}
H_{\text{eff}}^D&=&-J^{(2)}\sum_n\left( \hat{c}_{n}^{\dagger}\hat{c}_{n+1}+\text{h.c.}\right) \,\label{Eq:EffHam}\\
&+&\frac{U^{(2)}}{2}\sum_n \hat{N}_n^D\hat{N}_n^D
+V^{(2)}\sum_n \hat{N}_n^D\hat{N}_{n+1}^D \,,\nonumber
\end{eqnarray}
where $\hat{N}_n^D|N^D_n\rangle=\frac{\hat{n}_{n,a} + \hat{n}_{n,b}}{2}|N^D_n\rangle=N^D_n|N^D_n\rangle$ is the dimer number operator and $\hat{c}^{\dagger}_n$, $\hat{c}_n$ are the respective dimer creation and annihilation operators which satisfy $\left[ \hat{c}_n, \hat{c}^{\dagger}_m\right]=\delta_{n,m}$. The first term describes the hopping of the dimers with strength $J^{(2)}=2t^2(1+r)/U$. The second line describes on-site interactions between two dimers with strength $U^{(2)}=Ur-4t^2r/U$ and a nearest-neighbor interaction $V^{(2)}=-4t^2 (1-r)/U$. Notice that we included the cross terms proportional to $t^2r/U$ which play a major role in the formation of the liquid as we will show later.
\begin{figure}[t]
\includegraphics[width=1\columnwidth]{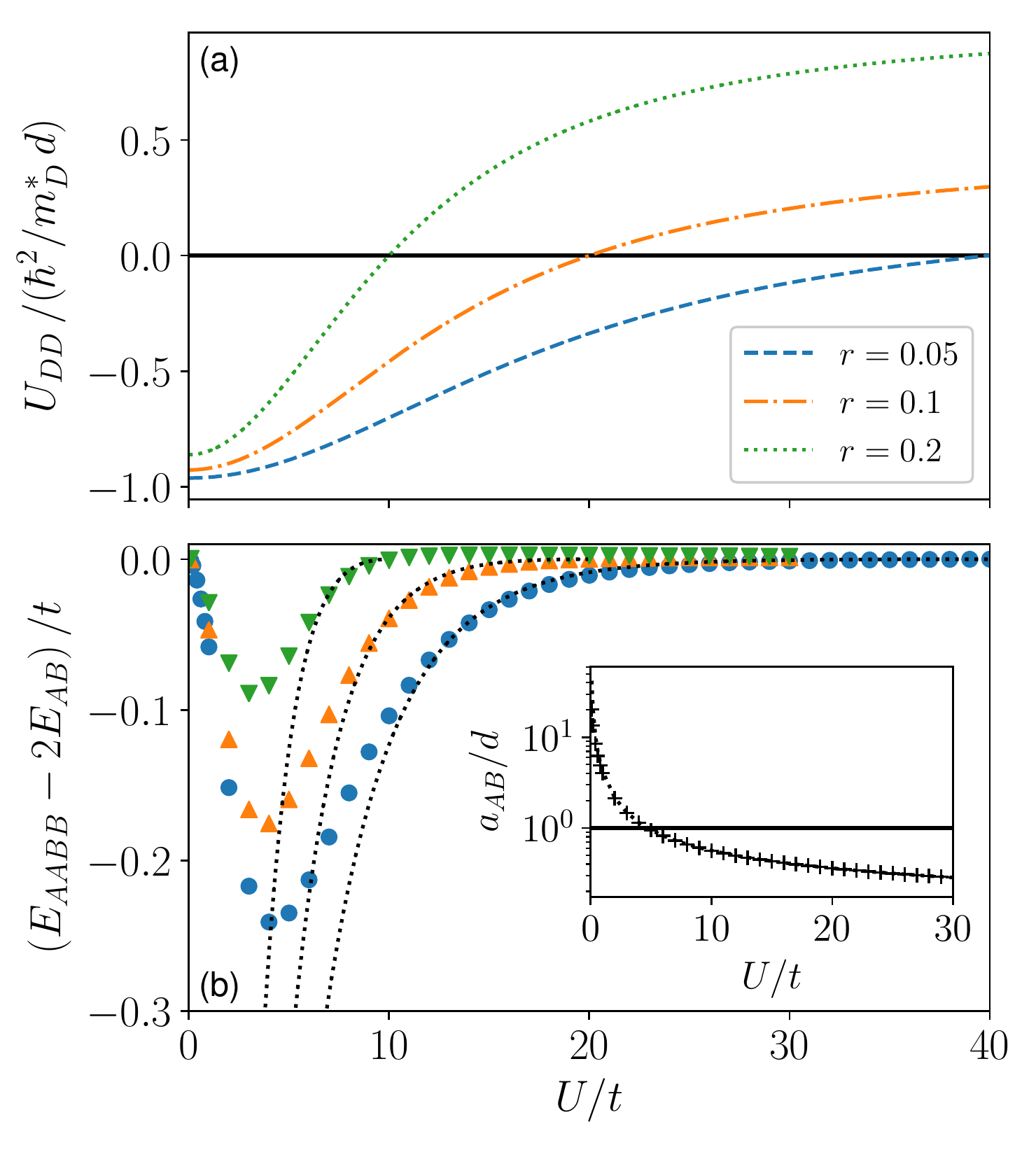} 
\caption{Panel (a): Strength of the effective dimer-dimer interaction as a function of the 
interaction $U/t$ for different values of the ratio $r$. Panel (b): Main figure: Tetramer energy as a function of the interaction $U/t$ for different ratios $r$. Analytical result (dotted lines) for the bound state energy using Eq.~\eqref{Eq:EnergyBound}. Inset: Typical length scale $a_{AB}$ associated with the two-particle bound state for $r=0.1$. Dotted line shows the analytical result for this length, see Eq.~\eqref{Eq:DimerSize}.}
\label{Fig:Scattering}
\end{figure}

The scattering problem of two particles described by the Hamiltonian~\eqref{Eq:EffHam} was solved in Ref.~\cite{Valiente_2009}. The resulting $s$-wave scattering length 
is given by~\cite{Valiente_2009},
$a_{\rm DD}/d=\left(U^{(2)}V^{(2)}-4J^{(2)}\left(2J^{(2)}- V^{(2)}\right)\right)/(U^{(2)}V^{(2)}+2J^{(2)}(2V^{(2)}+U^{(2)}))$ 
with $d$ the lattice spacing. In terms of characteristic parameters $U, t, r$ of the problem, we express it as
\begin{equation}
\frac{a_{\rm DD}}{d}=\frac{(r-1) r U^2/t^2-4 \left(r^2+3 r+4\right)}{2 r^2 U^2/t^2-8 }.
\label{Eq:scatt}
\end{equation}
Increasing the interaction $U/t$ for fixed $r$ we find a special point where the scattering length diverges and changes its sign, going from $a_{\rm DD}\rightarrow +\infty$ to $a_{\rm DD}\rightarrow -\infty$. The position of this threshold corresponds to a pole in $a_{DD}/d$, which results in the condition 
\begin{equation}
\label{Eq:threshold}
V^{(2)} = -2J^{(2)} \frac{U^{(2)}}{4J^{(2)}+U^{(2)}}\,\Rightarrow \, r_c=2 t/U\,.
\end{equation}
The effective one-dimensional dimer-dimer coupling constant, defined as $U_{\rm DD}=-2\hbar^2/(m^*_{\rm D} a_{\rm DD})$ with $m^*_{\rm D}=\hbar^2/(2J^{(2)}d^2)$ the effective mass of the dimer, crosses the zero value at the threshold. To the left (right) of this point, when $a_{\rm DD}>0$ ($a_{\rm DD}<0$) the effective dimer-dimer interactions are attractive (repulsive), see Fig.~\ref{Fig:Scattering}(a). In the attractive region a dimer-dimer bound state, i.e. tetramer, is formed and its energy vanishes when the threshold is reached. The binding energy of the tetramer in the vicinity of the threshold can be estimated by 
\begin{eqnarray}
E_B &\approx& -\frac{\hbar^2}{m^*_{\rm D} a_{\rm DD}^2}
=-\frac{1}{2 J^{(2)}}\left(\frac{2 J^{(2)} U^{(2)}}{4 J^{(2)}+U^{(2)}}+V^{(2)}\right)^2+\dots \nonumber\\
\label{Eq:EnergyBound}
\end{eqnarray}
Let us remark that the appearance of the threshold at a finite value of $U/t$ is a direct consequence of the cross terms proportional to $t^2r/U$ included in the effective Hamiltonian~\eqref{Eq:EffHam}. 

%
We resort to DMRG method to obtain the exact tetramer energy of the full Hamiltonian~\eqref{eq:TwoBH} in the four-particle $N_a=N_b=2$ and two-particle $N_a=N_b=1$ cases. In order to establish if a dimer-dimer bound state gets formed we compute $E_{AABB}-2E_{AB}$. Its negative value signals the formation of the bound state due to an effective attraction between dimers. Figure~\ref{Fig:Scattering}(b) shows the comparison of the exact tetramer energy with the analytical prediction given by the effective dimer model, Eq.~(\ref{Eq:EnergyBound}). An excellent agreement is found when $U/t\gg 1$ and $r\ll 1$, i.e. in the regime of deep dimers. To get a further insight, we calculate the dimer size $a_{AB}$ by associating it with the exact dimer energy $E_{AB}-2E_A = -2td^2/a_{AB}^2$ and alternatively with their asymptotic values for $L\rightarrow \infty$, $E_{AB}=-|U_{AB}|\sqrt{1+16 t^2/U_{AB}^2}$ and $E_A=-2t$~\cite{Valiente_2008}. This sets the relation,
\begin{eqnarray}
E_{AB}-2E_A&=&\frac{-2t}{\left(a_{AB}/d\right)^2}\,\label{Eq:DimerSize}\\
&=& -U|r-1|\sqrt{1+\frac{16t^2}{U^2(1-r)^2}}+4t\nonumber\,.
\end{eqnarray}
We compare the energy obtained in exact DMRG calculations and the asymptotic expression in the inset of Fig.~\ref{Fig:Scattering}(b). For $U/t\gg 1$ the dimer size is much smaller than the lattice spacing, $a_{AB}\ll d$. In this regime, it is possible to neglect the internal structure of dimers and treat them as composite bosons described by the effective Hamiltonian~\eqref{Eq:EffHam}. At the same time, the four-particle bound state might be large $a_{\rm DD}\gg d \gg a_{AB}$. Crucial differences appear in the opposite regime $U/t\ll 1$ where the effective composite boson model predicts a deeper bound state while exact results show that the bound state is shallower, see Fig.~\ref{Fig:Scattering}(b). In this regime we observe a dimer with an extension comparable to the four-body bound state $a_{\rm DD}\sim a_{AB}\gg d$, which makes the assumption of localized dimers no longer applicable.

%
\begin{figure}[t]
\includegraphics[width=1\columnwidth]{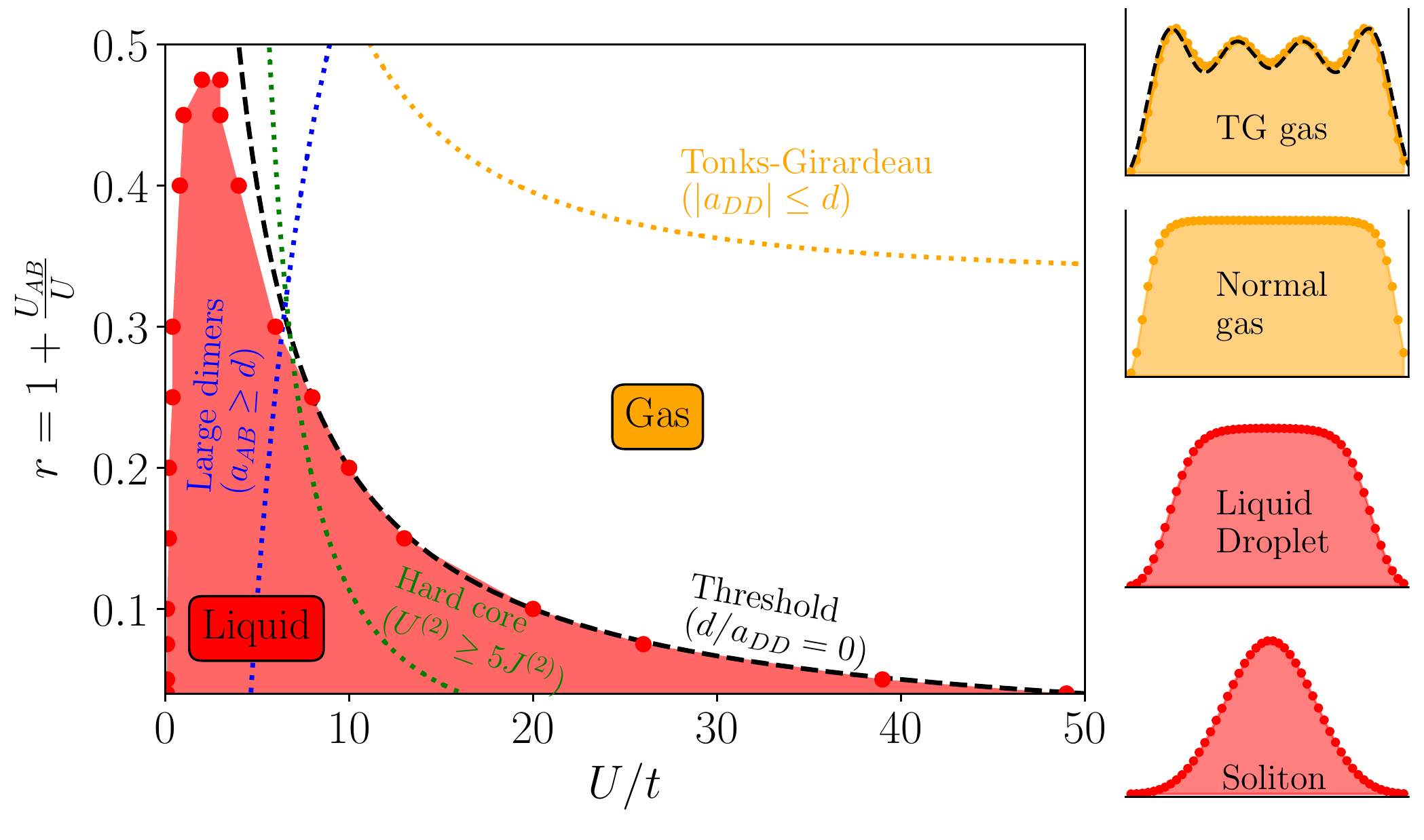} 
\caption{
Phase diagram in the plane of interaction strength $U/t$ and interaction imbalance $r$. The region where a tetramer bound state is formed is shown as a shaded (red) area. Its boundary, delimited by circles, is defined by a vanishing energy in the full Hamiltonian~\eqref{eq:TwoBH}, $E_{AABB}-2E_{AB}=0$. The boundary obtained within the effective dimer Hamiltonian is shown with a dashed line and corresponds to Eq.~\eqref{Eq:threshold}. The blue dotted line depicts the characteristic condition $a_{AB}=d$, e.g. to the left of the line, the dimer size is larger than the lattice spacing, see Eq.~\eqref{Eq:DimerSize}, and the effective Hamiltonian~\eqref{Eq:EffHam} does not apply. The green dotted line denotes the hard-core dimer condition $U^{(2)}\gg J^{(2)}$, to the right of this line the effective Hamiltonian~\eqref{Eq:HcHam} applies. To the right of the orange dotted line the Tonks-Girardeau regime is reached $|a_{DD}|\leq d$ and we recover the local properties of an ideal spinless fermionic gas.
In the many-body problem the different regimes still persist and the threshold line denotes the phase transition between a gas-liquid/soliton phase. On the right we show the typical dimer density profiles for the Tonks-Girardeau gas
together with the density of ideal spinless fermions (dashed line), 
the normal gas,
the liquid droplet, and the soliton. Note that the atom density profiles are very similar.}
\label{Fig:BoundRegion}
\end{figure}
The phase diagram in the $(r,U/t)$ plane is reported in Fig.~\ref{Fig:BoundRegion}. We find a sizeable region of parameters where a four-body bound state is formed. We discern two different regimes in the phase diagram separated by the $a_{AB}=d$ condition shown with a dotted line. To its right, the dimer size is smaller than the lattice spacing and the interactions are strong, $|U_{AB}|/t \approx U/t \ge10$. Here, the dimers are deeply bound and the effective dimer model is expected to be applicable. Indeed, it correctly predicts the boundary for tetramer formation, shown with a dashed line which is defined by a diverging dimer-dimer scattering length, $a_{\rm DD}\to \infty$, Eq.~\eqref{Eq:threshold}. In the second region, $U/t < 10$ the effective dimer model breaks down, as the dimers are no longer localized on a single lattice site. 
The tetramer bound state completely disappears for interaction imbalance larger than $r\approx 0.475$, a slightly smaller value than $r\approx 0.53$ reported in the continuum~\cite{PhysRevA.97.063616,PhysRevLett.122.105302}.

Once the threshold line is crossed, $a_{\rm DD}<0$, the effective dimer-dimer interaction becomes repulsive and tetramer formation does not happen. At the same time the dimers are still formed and repulsion between them becomes stronger as $U/t$ is increased. 
Eventually, when $|a_{\rm DD}|\ll d$, the Tonks-Girardeau regime is reached and the dimers fermionize. The strong repulsion between dimers does not allow them to stay at the same lattice site, mimicking the Fermi exclusion principle. As a result, the energetic and local properties of dimers are expected to be similar to those of ideal fermions. In order to demonstrate that we compute the dimer density profile $n_i^D=\langle \hat{d}_i^{\dagger}\hat{d}_i \rangle$ with the bosonic dimer operator $\hat{d}_i=\hat{a}_i\hat{b}_i$ for different system with $N_A=N_B=N/2$ particles. The profiles obtained are almost equal to the ones corresponding to a system of ideal fermions, see Fig.~\ref{Fig:BoundRegion} for an example with $N=8$.

{\bf Many-dimer problem.}
\begin{figure}[t]
\includegraphics[width=1\columnwidth]{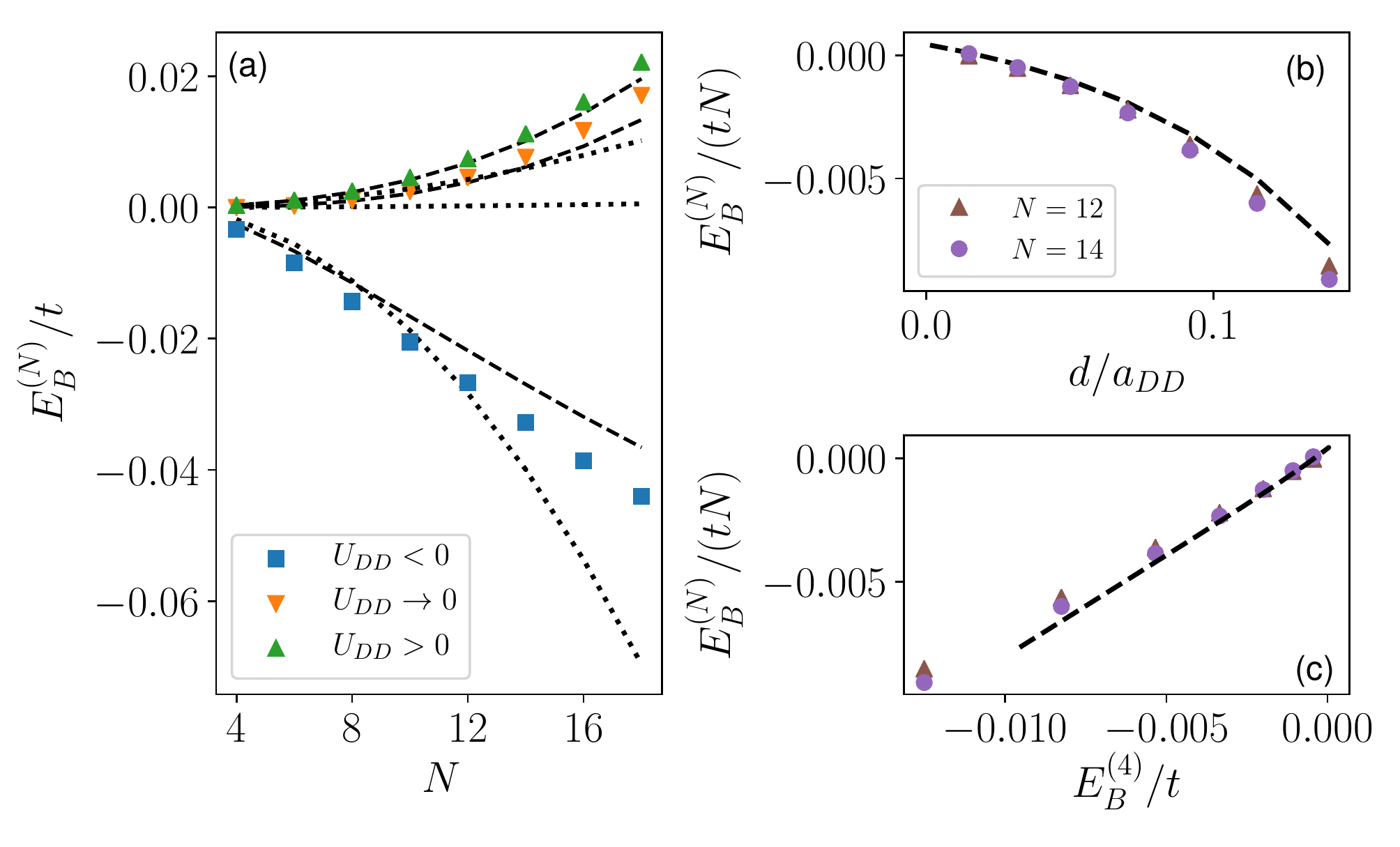}
\caption{Panel (a): Energy of the $N$-body state $E_B^{(N)}=E^{(N)}-NE^{(2)}/2$ as a function of the total number of particles $N$ for fixed ratio $r=0.1$ and three characteristic interaction strengths, $U=16, 19.95, 21$, corresponding to $U_{\rm DD}<0$, $U_{\rm DD}\simeq 0$ and $U_{\rm DD}>0$, respectively. The results are obtained for the full Hamiltonian~\eqref{eq:TwoBH} (symbols), for the effective model of composite bosons matching the dimer-dimer scattering length (dotted line) and for the full effective dimer model, Eq.~(\ref{Eq:EffHam}) (dashed line). Panels~(b) and~(c): Energy of the $N$-body state per particle for $U_{\rm DD}<0$ as a function of the inverse of the dimer-dimer scattering length (b) or the dimer-dimer bound energy (c) for the full Hamiltonian (symbols) and the effective dimer model~\eqref{Eq:EffHam} (dashed line).}
\label{Fig:EnergyN}
\end{figure}
As we have shown, in the regime of deeply bound dimers, the four-body problem can be interpreted in terms of composite bosons interacting via an effective potential. Here we show that a similar interpretation holds for any number of atoms. To this end, we compute the binding energy of the $N$-body  state $E_B^{(N)}=E^{(N)}-NE^{(2)}/2$ for the case where the dimer-dimer effective interaction is (a) attractive, (b) vanishing, and (c) repulsive. 

For repulsive effective dimer-dimer interactions, $U_{\rm DD}>0$, the energy of $N$-body state is positive and grows nearly quadratically with the number of particles, which can be interpreted as a linear increase of the chemical potential of a gas as the density is augmented, see Fig.~\ref{Fig:EnergyN}(a). In the attractive dimer-dimer case, $U_{\rm DD}<0$, we observe a nearly constant negative energy per particle which signals the presence of a stable liquid phase, as shown in Fig.~\ref{Fig:EnergyN}(a). This physics is fully captured by the effective dimer-dimer Hamiltonian of Eq.~(\ref{Eq:EffHam}) as can be seen by comparing the dashed lines with the symbols in Fig.~\ref{Fig:EnergyN}(a). The formation of a liquid in the attractive dimer-dimer regime is not trivial, as in principle the bosonic dimers could form a soliton and eventually condense on a single site. In the continuum, the interpretation of the stability provided in Refs.~\cite{PhysRevLett.89.050402,PhysRevA.97.063616} involves introduction of microscopic repulsive three-dimer interactions which counterbalance the dimer-dimer attraction. In our case, the effective dimer model correctly captures the formation of the liquid and thus it must incorporate a stabilization mechanism. Indeed we find that in our case the liquid is stabilized by the non-zero effective range of the dimer-dimer interaction stemming from Eq.~(\ref{Eq:EffHam}). Close to the threshold, the effective range~\cite{Valiente_2009} 
\begin{equation}
\frac{r_e}{d} \simeq -\frac{V^{(2)}}{4J^{(2)}}\left (4+\frac{V^{(2)}}{J^{(2)}}\right) = \frac{(1-r_c) (3 r_c+1)}{(r_c+1)^2},
\end{equation}
changes smoothly contrarily to the resonant behavior of the effective dimer-dimer scattering length $a_{\rm DD}$. Notice that in this regime, the effective range is proportional to the nearest-neighbor interaction $V^{(2)}$ and is of the order of the lattice spacing $r_e/d\approx 1-4r_c^2$ for $r_c\ll 1$. To elucidate the role of the effective range we perform calculations using a dimer-dimer model without the $V^{(2)}$ term, and with the effective interaction $U^{(2)}$ tuned such that the model reproduces the same scattering length as the full dimer-dimer Hamiltonian. In this case, the energy per particle is not constant and becomes more negative for increasing the number of particles, going from the formation of a McGuire soliton~\cite{McGuire64} to the collapse of all dimers at one site~\cite{SCOTT1994194}, see dotted line in Fig.~\ref{Fig:EnergyN}(a). 
The McGuire soliton has a cubic dependence of the energy on the number of $N_D$~\cite{McGuire64}, $E = -N_D(N_D^2-1)2\hbar^2/(3m_D^*a_{\rm DD}^2)$ 
and is sustained for sufficiently small number of particles, such that its size is large compared to the lattice spacing, $a_{\rm DD}/N_D \gg d$. On the other hand, when the size of the soliton becomes of the order of the lattice spacing $a_{\rm DD}/N_D \sim d$ the energy exhibits a quadratic dependence on $N_D$, $E\propto U_{DD}N_D(N_D-1)/2$, see Ref.~\cite{SCOTT1994194}. The dotted line for attractive dimer-dimer interaction in Fig.~\ref{Fig:EnergyN}(a) follows this behavior.

In addition, we verify that the effective dimer-dimer model describes correctly the energy of the liquid as a function of the interaction strength reported in Fig.~\ref{Fig:EnergyN}(b) for a fixed number of particles. We observe that close to the threshold, the energy of the $N$-body problem is linearly proportional to the energy of the two-dimer problem as shown in Fig.~\ref{Fig:EnergyN}(c). This linear relation suggests a dimerized and universal nature of the self-bound liquid, since the properties of the $N$-body liquid can be directly expressed in terms of the dimer-dimer energy. 

%
%

The liquid is very dilute in the vicinity of the threshold of its appearance. In particular, the probability of finding two dimers in the same site is extremely small. This suggests that an effective description in terms of hard-core dimers with a finite-range attraction could explain the liquid formation in the full Hamiltonian~\eqref{eq:TwoBH}. The hard-core description applies when the local dimer effective interaction is much stronger than the effective hopping $U^{(2)}\gg J^{(2)}$ which leads to $(U/t)^2r\gg1$. Therefore, there is a window in the regime of validity of the effective dimer Hamiltonian $U/t\gg1$ and $r\ll1$ where the hard-core condition is satisfied. In order to impose the hard-core constrain on the dimers ($\left(\hat{c}^{\dagger}\right)^2|0\rangle = 0$) we write a new effective Hamiltonian at second order in $J^{(2)}$~\cite{PhysRevA.67.053606,PhysRevB.87.035104},
\begin{eqnarray}
H_{\text{eff}}^{\rm HCD}&=&-J^{(2)}_{\rm HC}\sum_n\!\left( 
\hat{c}_{n}^{\dagger}\hat{c}_{n+1}+\text{h.c.}\right) \nonumber \\
&+&V^{(2)}_{\rm HC}\sum_n\!\hat{N}_n^D\hat{N}_{n+1}^D  \,,
\label{Eq:HcHam}
\end{eqnarray}
with $J^{(2)}_{\rm HC}\approx J^{(2)}\left(1+ \left(4J^{(2)}/U^{(2)} \right)^2 \right)$ and $V^{(2)}_{\rm HC}=V^{(2)}-8\left(J^{(2)}\right)^2/U^{(2)}$. This hard-core model has been extensively studied and it presents a phase transition at $V^{(2)}_{\rm HC}=-2J^{(2)}_{\rm HC}$~\cite{RevModPhys.83.1405,Giamarchi:743140} which leads to the condition,
\begin{equation}
V^{(2)} = -2J^{(2)} \frac{U^{(2)}}{4J^{(2)}+U^{(2)}}.
\end{equation}
Strikingly, this is nothing else but the condition of finding a pole in the effective dimer-dimer scattering length Eq.~\eqref{Eq:threshold}. Therefore we conclude that the hard-core effective dimer model can explain the liquid formation of the complete Hamiltonian~\eqref{eq:TwoBH}. First of all, the strong on-site repulsion $U^{(2)}$ avoids the collapse of the system. 
Then the attractive nearest-neighbor interaction, which sets the effective range of the dimer-dimer scattering problem, bounds the system and stabilizes the liquid phase. 

{\bf Conclusions} 
We have shown that a bosonic mixture trapped in a one-dimensional lattice with attractive interspecies and repulsive intraspecies interactions has a parameter region where liquid, gas and soliton phases appear. Studying the four-atom problem, we derive an effective dimer-dimer Hamiltonian which correctly describes the gas-liquid (or soliton) phase transition when the system is formed by deeply bound dimers. This phase transition is marked by the resonance of the dimer-dimer scattering length. Moreover, the liquid state exhibits a universal behavior since only depends on the parameters of the two-dimer scattering problem, namely the scattering length and the effective range.
These ingredients 
are enough to predict the existence of self-bound objects. 
This has to be compared with the stabilization mechanism used in the continuum counterpart where a three-body repulsion is needed in order to stabilize the liquid~\cite{PhysRevLett.89.050402,PhysRevA.97.063616}. 
In the vicinity of the resonance, the liquid is very dilute and can be described in terms of hard-core dimers with an attractive nearest-neighbor interaction. For strong repulsive dimer-dimer interactions bosonic dimers experience fermionization thus reaching the Tonks-Girardeau regime. The applicability of our results goes beyond the study of bosonic mixtures and the universal liquid phase could also be observed in other systems such as dipolar bosons in optical lattices~\cite{Lahaye_2009,Trefzger_2011}. Specifically, similar Hamiltonians appear in bilayer optical lattices~\cite{PhysRevA.75.053613,PhysRevLett.103.035304} 
In addition the predicted phases are directly accessible with current techniques used in ultracold quantum gases laboratories~\cite{Schmitt2016,Cabrera301,PhysRevResearch.1.033155,2020arXiv200509549J}.

{\bf Acknowledgements.} 
This work has been partially supported by MINECO (Spain) Grants No. FIS2017- 87534-P and FIS2017-84114-C2-1-P and by the European Union Regional Development Fund within the ERDF Operational Program of Catalunya (project QUASICAT/QuantumCat). DRMG computations have been performed using TeNPy~\cite{10.21468/SciPostPhysLectNotes.5}.

\bibliographystyle{apsrev4-1}
\bibliography{paperbib}

\begin{thebibliography}{40}%
\makeatletter
\providecommand \@ifxundefined [1]{%
 \@ifx{#1\undefined}
}%
\providecommand \@ifnum [1]{%
 \ifnum #1\expandafter \@firstoftwo
 \else \expandafter \@secondoftwo
 \fi
}%
\providecommand \@ifx [1]{%
 \ifx #1\expandafter \@firstoftwo
 \else \expandafter \@secondoftwo
 \fi
}%
\providecommand \natexlab [1]{#1}%
\providecommand \enquote  [1]{``#1''}%
\providecommand \bibnamefont  [1]{#1}%
\providecommand \bibfnamefont [1]{#1}%
\providecommand \citenamefont [1]{#1}%
\providecommand \href@noop [0]{\@secondoftwo}%
\providecommand \href [0]{\begingroup \@sanitize@url \@href}%
\providecommand \@href[1]{\@@startlink{#1}\@@href}%
\providecommand \@@href[1]{\endgroup#1\@@endlink}%
\providecommand \@sanitize@url [0]{\catcode `\\12\catcode `\$12\catcode
  `\&12\catcode `\#12\catcode `\^12\catcode `\_12\catcode `\%12\relax}%
\providecommand \@@startlink[1]{}%
\providecommand \@@endlink[0]{}%
\providecommand \url  [0]{\begingroup\@sanitize@url \@url }%
\providecommand \@url [1]{\endgroup\@href {#1}{\urlprefix }}%
\providecommand \urlprefix  [0]{URL }%
\providecommand \Eprint [0]{\href }%
\providecommand \doibase [0]{http://dx.doi.org/}%
\providecommand \selectlanguage [0]{\@gobble}%
\providecommand \bibinfo  [0]{\@secondoftwo}%
\providecommand \bibfield  [0]{\@secondoftwo}%
\providecommand \translation [1]{[#1]}%
\providecommand \BibitemOpen [0]{}%
\providecommand \bibitemStop [0]{}%
\providecommand \bibitemNoStop [0]{.\EOS\space}%
\providecommand \EOS [0]{\spacefactor3000\relax}%
\providecommand \BibitemShut  [1]{\csname bibitem#1\endcsname}%
\let\auto@bib@innerbib\@empty
\bibitem [{\citenamefont {Hansen}\ and\ \citenamefont
  {McDonald}(2013)}]{hansen}%
  \BibitemOpen
  \bibfield  {author} {\bibinfo {author} {\bibfnamefont {J.~P.}\ \bibnamefont
  {Hansen}}\ and\ \bibinfo {author} {\bibfnamefont {I.~R.}\ \bibnamefont
  {McDonald}},\ }\href@noop {} {\emph {\bibinfo {title} {Theory of Simple
  Liquids}}}\ (\bibinfo  {publisher} {Academic Press},\ \bibinfo {address}
  {Oxford},\ \bibinfo {year} {2013})\BibitemShut {NoStop}%
\bibitem [{\citenamefont {Fetter}\ and\ \citenamefont
  {Walecka}(2003)}]{fetter2003quantum}%
  \BibitemOpen
  \bibfield  {author} {\bibinfo {author} {\bibfnamefont {A.}~\bibnamefont
  {Fetter}}\ and\ \bibinfo {author} {\bibfnamefont {J.}~\bibnamefont
  {Walecka}},\ }\href {https://books.google.es/books?id=0wekf1s83b0C} {\emph
  {\bibinfo {title} {Quantum Theory of Many-particle Systems}}},\ Dover Books
  on Physics\ (\bibinfo  {publisher} {Dover Publications},\ \bibinfo {year}
  {2003})\BibitemShut {NoStop}%
\bibitem [{\citenamefont {Leggett}(2006)}]{leggett2006quantum}%
  \BibitemOpen
  \bibfield  {author} {\bibinfo {author} {\bibfnamefont {A.}~\bibnamefont
  {Leggett}},\ }\href {https://books.google.es/books?id=PiRRAAAAMAAJ} {\emph
  {\bibinfo {title} {Quantum Liquids: Bose Condensation and Cooper Pairing in
  Condensed-matter Systems}}},\ Oxford graduate texts in mathematics\ (\bibinfo
   {publisher} {OUP Oxford},\ \bibinfo {year} {2006})\BibitemShut {NoStop}%
\bibitem [{\citenamefont {Ferrier-Barbut}\ \emph {et~al.}(2016)\citenamefont
  {Ferrier-Barbut}, \citenamefont {Kadau}, \citenamefont {Schmitt},
  \citenamefont {Wenzel},\ and\ \citenamefont {Pfau}}]{PhysRevLett.116.215301}%
  \BibitemOpen
  \bibfield  {author} {\bibinfo {author} {\bibfnamefont {I.}~\bibnamefont
  {Ferrier-Barbut}}, \bibinfo {author} {\bibfnamefont {H.}~\bibnamefont
  {Kadau}}, \bibinfo {author} {\bibfnamefont {M.}~\bibnamefont {Schmitt}},
  \bibinfo {author} {\bibfnamefont {M.}~\bibnamefont {Wenzel}}, \ and\ \bibinfo
  {author} {\bibfnamefont {T.}~\bibnamefont {Pfau}},\ }\href {\doibase
  10.1103/PhysRevLett.116.215301} {\bibfield  {journal} {\bibinfo  {journal}
  {Phys. Rev. Lett.}\ }\textbf {\bibinfo {volume} {116}},\ \bibinfo {pages}
  {215301} (\bibinfo {year} {2016})}\BibitemShut {NoStop}%
\bibitem [{\citenamefont {Schmitt}\ \emph {et~al.}(2016)\citenamefont
  {Schmitt}, \citenamefont {Wenzel}, \citenamefont {B{\"o}ttcher},
  \citenamefont {Ferrier-Barbut},\ and\ \citenamefont {Pfau}}]{Schmitt2016}%
  \BibitemOpen
  \bibfield  {author} {\bibinfo {author} {\bibfnamefont {M.}~\bibnamefont
  {Schmitt}}, \bibinfo {author} {\bibfnamefont {M.}~\bibnamefont {Wenzel}},
  \bibinfo {author} {\bibfnamefont {F.}~\bibnamefont {B{\"o}ttcher}}, \bibinfo
  {author} {\bibfnamefont {I.}~\bibnamefont {Ferrier-Barbut}}, \ and\ \bibinfo
  {author} {\bibfnamefont {T.}~\bibnamefont {Pfau}},\ }\href {\doibase
  10.1038/nature20126} {\bibfield  {journal} {\bibinfo  {journal} {Nature}\
  }\textbf {\bibinfo {volume} {539}},\ \bibinfo {pages} {259} (\bibinfo {year}
  {2016})}\BibitemShut {NoStop}%
\bibitem [{\citenamefont {Chomaz}\ \emph {et~al.}(2016)\citenamefont {Chomaz},
  \citenamefont {Baier}, \citenamefont {Petter}, \citenamefont {Mark},
  \citenamefont {W\"achtler}, \citenamefont {Santos},\ and\ \citenamefont
  {Ferlaino}}]{PhysRevX.6.041039}%
  \BibitemOpen
  \bibfield  {author} {\bibinfo {author} {\bibfnamefont {L.}~\bibnamefont
  {Chomaz}}, \bibinfo {author} {\bibfnamefont {S.}~\bibnamefont {Baier}},
  \bibinfo {author} {\bibfnamefont {D.}~\bibnamefont {Petter}}, \bibinfo
  {author} {\bibfnamefont {M.~J.}\ \bibnamefont {Mark}}, \bibinfo {author}
  {\bibfnamefont {F.}~\bibnamefont {W\"achtler}}, \bibinfo {author}
  {\bibfnamefont {L.}~\bibnamefont {Santos}}, \ and\ \bibinfo {author}
  {\bibfnamefont {F.}~\bibnamefont {Ferlaino}},\ }\href {\doibase
  10.1103/PhysRevX.6.041039} {\bibfield  {journal} {\bibinfo  {journal} {Phys.
  Rev. X}\ }\textbf {\bibinfo {volume} {6}},\ \bibinfo {pages} {041039}
  (\bibinfo {year} {2016})}\BibitemShut {NoStop}%
\bibitem [{\citenamefont {Cabrera}\ \emph {et~al.}(2018)\citenamefont
  {Cabrera}, \citenamefont {Tanzi}, \citenamefont {Sanz}, \citenamefont
  {Naylor}, \citenamefont {Thomas}, \citenamefont {Cheiney},\ and\
  \citenamefont {Tarruell}}]{Cabrera301}%
  \BibitemOpen
  \bibfield  {author} {\bibinfo {author} {\bibfnamefont {C.~R.}\ \bibnamefont
  {Cabrera}}, \bibinfo {author} {\bibfnamefont {L.}~\bibnamefont {Tanzi}},
  \bibinfo {author} {\bibfnamefont {J.}~\bibnamefont {Sanz}}, \bibinfo {author}
  {\bibfnamefont {B.}~\bibnamefont {Naylor}}, \bibinfo {author} {\bibfnamefont
  {P.}~\bibnamefont {Thomas}}, \bibinfo {author} {\bibfnamefont
  {P.}~\bibnamefont {Cheiney}}, \ and\ \bibinfo {author} {\bibfnamefont
  {L.}~\bibnamefont {Tarruell}},\ }\href {\doibase 10.1126/science.aao5686}
  {\bibfield  {journal} {\bibinfo  {journal} {Science}\ }\textbf {\bibinfo
  {volume} {359}},\ \bibinfo {pages} {301} (\bibinfo {year}
  {2018})}\BibitemShut {NoStop}%
\bibitem [{\citenamefont {Cheiney}\ \emph {et~al.}(2018)\citenamefont
  {Cheiney}, \citenamefont {Cabrera}, \citenamefont {Sanz}, \citenamefont
  {Naylor}, \citenamefont {Tanzi},\ and\ \citenamefont
  {Tarruell}}]{PhysRevLett.120.135301}%
  \BibitemOpen
  \bibfield  {author} {\bibinfo {author} {\bibfnamefont {P.}~\bibnamefont
  {Cheiney}}, \bibinfo {author} {\bibfnamefont {C.~R.}\ \bibnamefont
  {Cabrera}}, \bibinfo {author} {\bibfnamefont {J.}~\bibnamefont {Sanz}},
  \bibinfo {author} {\bibfnamefont {B.}~\bibnamefont {Naylor}}, \bibinfo
  {author} {\bibfnamefont {L.}~\bibnamefont {Tanzi}}, \ and\ \bibinfo {author}
  {\bibfnamefont {L.}~\bibnamefont {Tarruell}},\ }\href {\doibase
  10.1103/PhysRevLett.120.135301} {\bibfield  {journal} {\bibinfo  {journal}
  {Phys. Rev. Lett.}\ }\textbf {\bibinfo {volume} {120}},\ \bibinfo {pages}
  {135301} (\bibinfo {year} {2018})}\BibitemShut {NoStop}%
\bibitem [{\citenamefont {Semeghini}\ \emph {et~al.}(2018)\citenamefont
  {Semeghini}, \citenamefont {Ferioli}, \citenamefont {Masi}, \citenamefont
  {Mazzinghi}, \citenamefont {Wolswijk}, \citenamefont {Minardi}, \citenamefont
  {Modugno}, \citenamefont {Modugno}, \citenamefont {Inguscio},\ and\
  \citenamefont {Fattori}}]{PhysRevLett.120.235301}%
  \BibitemOpen
  \bibfield  {author} {\bibinfo {author} {\bibfnamefont {G.}~\bibnamefont
  {Semeghini}}, \bibinfo {author} {\bibfnamefont {G.}~\bibnamefont {Ferioli}},
  \bibinfo {author} {\bibfnamefont {L.}~\bibnamefont {Masi}}, \bibinfo {author}
  {\bibfnamefont {C.}~\bibnamefont {Mazzinghi}}, \bibinfo {author}
  {\bibfnamefont {L.}~\bibnamefont {Wolswijk}}, \bibinfo {author}
  {\bibfnamefont {F.}~\bibnamefont {Minardi}}, \bibinfo {author} {\bibfnamefont
  {M.}~\bibnamefont {Modugno}}, \bibinfo {author} {\bibfnamefont
  {G.}~\bibnamefont {Modugno}}, \bibinfo {author} {\bibfnamefont
  {M.}~\bibnamefont {Inguscio}}, \ and\ \bibinfo {author} {\bibfnamefont
  {M.}~\bibnamefont {Fattori}},\ }\href {\doibase
  10.1103/PhysRevLett.120.235301} {\bibfield  {journal} {\bibinfo  {journal}
  {Phys. Rev. Lett.}\ }\textbf {\bibinfo {volume} {120}},\ \bibinfo {pages}
  {235301} (\bibinfo {year} {2018})}\BibitemShut {NoStop}%
\bibitem [{\citenamefont {D'Errico}\ \emph {et~al.}(2019)\citenamefont
  {D'Errico}, \citenamefont {Burchianti}, \citenamefont {Prevedelli},
  \citenamefont {Salasnich}, \citenamefont {Ancilotto}, \citenamefont
  {Modugno}, \citenamefont {Minardi},\ and\ \citenamefont
  {Fort}}]{PhysRevResearch.1.033155}%
  \BibitemOpen
  \bibfield  {author} {\bibinfo {author} {\bibfnamefont {C.}~\bibnamefont
  {D'Errico}}, \bibinfo {author} {\bibfnamefont {A.}~\bibnamefont
  {Burchianti}}, \bibinfo {author} {\bibfnamefont {M.}~\bibnamefont
  {Prevedelli}}, \bibinfo {author} {\bibfnamefont {L.}~\bibnamefont
  {Salasnich}}, \bibinfo {author} {\bibfnamefont {F.}~\bibnamefont
  {Ancilotto}}, \bibinfo {author} {\bibfnamefont {M.}~\bibnamefont {Modugno}},
  \bibinfo {author} {\bibfnamefont {F.}~\bibnamefont {Minardi}}, \ and\
  \bibinfo {author} {\bibfnamefont {C.}~\bibnamefont {Fort}},\ }\href {\doibase
  10.1103/PhysRevResearch.1.033155} {\bibfield  {journal} {\bibinfo  {journal}
  {Phys. Rev. Research}\ }\textbf {\bibinfo {volume} {1}},\ \bibinfo {pages}
  {033155} (\bibinfo {year} {2019})}\BibitemShut {NoStop}%
\bibitem [{\citenamefont {Petrov}(2015)}]{PhysRevLett.115.155302}%
  \BibitemOpen
  \bibfield  {author} {\bibinfo {author} {\bibfnamefont {D.~S.}\ \bibnamefont
  {Petrov}},\ }\href {\doibase 10.1103/PhysRevLett.115.155302} {\bibfield
  {journal} {\bibinfo  {journal} {Phys. Rev. Lett.}\ }\textbf {\bibinfo
  {volume} {115}},\ \bibinfo {pages} {155302} (\bibinfo {year}
  {2015})}\BibitemShut {NoStop}%
\bibitem [{\citenamefont {Baillie}\ \emph {et~al.}(2016)\citenamefont
  {Baillie}, \citenamefont {Wilson}, \citenamefont {Bisset},\ and\
  \citenamefont {Blakie}}]{PhysRevA.94.021602}%
  \BibitemOpen
  \bibfield  {author} {\bibinfo {author} {\bibfnamefont {D.}~\bibnamefont
  {Baillie}}, \bibinfo {author} {\bibfnamefont {R.~M.}\ \bibnamefont {Wilson}},
  \bibinfo {author} {\bibfnamefont {R.~N.}\ \bibnamefont {Bisset}}, \ and\
  \bibinfo {author} {\bibfnamefont {P.~B.}\ \bibnamefont {Blakie}},\ }\href
  {\doibase 10.1103/PhysRevA.94.021602} {\bibfield  {journal} {\bibinfo
  {journal} {Phys. Rev. A}\ }\textbf {\bibinfo {volume} {94}},\ \bibinfo
  {pages} {021602} (\bibinfo {year} {2016})}\BibitemShut {NoStop}%
\bibitem [{\citenamefont {W\"achtler}\ and\ \citenamefont
  {Santos}(2016)}]{PhysRevA.94.043618}%
  \BibitemOpen
  \bibfield  {author} {\bibinfo {author} {\bibfnamefont {F.}~\bibnamefont
  {W\"achtler}}\ and\ \bibinfo {author} {\bibfnamefont {L.}~\bibnamefont
  {Santos}},\ }\href {\doibase 10.1103/PhysRevA.94.043618} {\bibfield
  {journal} {\bibinfo  {journal} {Phys. Rev. A}\ }\textbf {\bibinfo {volume}
  {94}},\ \bibinfo {pages} {043618} (\bibinfo {year} {2016})}\BibitemShut
  {NoStop}%
\bibitem [{\citenamefont {Kartashov}\ \emph {et~al.}(2019)\citenamefont
  {Kartashov}, \citenamefont {Astrakharchik}, \citenamefont {Malomed},\ and\
  \citenamefont {Torner}}]{Kartashov2019}%
  \BibitemOpen
  \bibfield  {author} {\bibinfo {author} {\bibfnamefont {Y.~V.}\ \bibnamefont
  {Kartashov}}, \bibinfo {author} {\bibfnamefont {G.~E.}\ \bibnamefont
  {Astrakharchik}}, \bibinfo {author} {\bibfnamefont {B.~A.}\ \bibnamefont
  {Malomed}}, \ and\ \bibinfo {author} {\bibfnamefont {L.}~\bibnamefont
  {Torner}},\ }\href {\doibase 10.1038/s42254-019-0025-7} {\bibfield  {journal}
  {\bibinfo  {journal} {Nat. Rev. Phys.}\ }\textbf {\bibinfo {volume} {1}},\
  \bibinfo {pages} {185} (\bibinfo {year} {2019})}\BibitemShut {NoStop}%
\bibitem [{\citenamefont {Petrov}\ and\ \citenamefont
  {Astrakharchik}(2016)}]{PhysRevLett.117.100401}%
  \BibitemOpen
  \bibfield  {author} {\bibinfo {author} {\bibfnamefont {D.~S.}\ \bibnamefont
  {Petrov}}\ and\ \bibinfo {author} {\bibfnamefont {G.~E.}\ \bibnamefont
  {Astrakharchik}},\ }\href {\doibase 10.1103/PhysRevLett.117.100401}
  {\bibfield  {journal} {\bibinfo  {journal} {Phys. Rev. Lett.}\ }\textbf
  {\bibinfo {volume} {117}},\ \bibinfo {pages} {100401} (\bibinfo {year}
  {2016})}\BibitemShut {NoStop}%
\bibitem [{\citenamefont {Parisi}\ \emph {et~al.}(2019)\citenamefont {Parisi},
  \citenamefont {Astrakharchik},\ and\ \citenamefont
  {Giorgini}}]{PhysRevLett.122.105302}%
  \BibitemOpen
  \bibfield  {author} {\bibinfo {author} {\bibfnamefont {L.}~\bibnamefont
  {Parisi}}, \bibinfo {author} {\bibfnamefont {G.~E.}\ \bibnamefont
  {Astrakharchik}}, \ and\ \bibinfo {author} {\bibfnamefont {S.}~\bibnamefont
  {Giorgini}},\ }\href {\doibase 10.1103/PhysRevLett.122.105302} {\bibfield
  {journal} {\bibinfo  {journal} {Phys. Rev. Lett.}\ }\textbf {\bibinfo
  {volume} {122}},\ \bibinfo {pages} {105302} (\bibinfo {year}
  {2019})}\BibitemShut {NoStop}%
\bibitem [{\citenamefont {Parisi}\ and\ \citenamefont
  {Giorgini}(2020)}]{ParisiGiorgini2020}%
  \BibitemOpen
  \bibfield  {author} {\bibinfo {author} {\bibfnamefont {L.}~\bibnamefont
  {Parisi}}\ and\ \bibinfo {author} {\bibfnamefont {S.}~\bibnamefont
  {Giorgini}},\ }\href@noop {} {} (\bibinfo {year} {2020}),\ \Eprint
  {http://arxiv.org/abs/arXiv:2003.05231} {arXiv:2003.05231} \BibitemShut
  {NoStop}%
\bibitem [{\citenamefont {Ota}\ and\ \citenamefont
  {Astrakharchik}(2020)}]{Ota2020}%
  \BibitemOpen
  \bibfield  {author} {\bibinfo {author} {\bibfnamefont {M.}~\bibnamefont
  {Ota}}\ and\ \bibinfo {author} {\bibfnamefont {G.~E.}\ \bibnamefont
  {Astrakharchik}},\ }\href@noop {} {} (\bibinfo {year} {2020}),\ \Eprint
  {http://arxiv.org/abs/arXiv:2005.10047} {arXiv:2005.10047} \BibitemShut
  {NoStop}%
\bibitem [{\citenamefont {Morera}\ \emph {et~al.}(2020)\citenamefont {Morera},
  \citenamefont {Astrakharchik}, \citenamefont {Polls},\ and\ \citenamefont
  {Juli\'a-D\'{\i}az}}]{PhysRevResearch.2.022008}%
  \BibitemOpen
  \bibfield  {author} {\bibinfo {author} {\bibfnamefont {I.}~\bibnamefont
  {Morera}}, \bibinfo {author} {\bibfnamefont {G.~E.}\ \bibnamefont
  {Astrakharchik}}, \bibinfo {author} {\bibfnamefont {A.}~\bibnamefont
  {Polls}}, \ and\ \bibinfo {author} {\bibfnamefont {B.}~\bibnamefont
  {Juli\'a-D\'{\i}az}},\ }\href {\doibase 10.1103/PhysRevResearch.2.022008}
  {\bibfield  {journal} {\bibinfo  {journal} {Phys. Rev. Research}\ }\textbf
  {\bibinfo {volume} {2}},\ \bibinfo {pages} {022008} (\bibinfo {year}
  {2020})}\BibitemShut {NoStop}%
\bibitem [{\citenamefont {Pricoupenko}\ and\ \citenamefont
  {Petrov}(2018)}]{PhysRevA.97.063616}%
  \BibitemOpen
  \bibfield  {author} {\bibinfo {author} {\bibfnamefont {A.}~\bibnamefont
  {Pricoupenko}}\ and\ \bibinfo {author} {\bibfnamefont {D.~S.}\ \bibnamefont
  {Petrov}},\ }\href {\doibase 10.1103/PhysRevA.97.063616} {\bibfield
  {journal} {\bibinfo  {journal} {Phys. Rev. A}\ }\textbf {\bibinfo {volume}
  {97}},\ \bibinfo {pages} {063616} (\bibinfo {year} {2018})}\BibitemShut
  {NoStop}%
\bibitem [{\citenamefont {Guijarro}\ \emph {et~al.}(2018)\citenamefont
  {Guijarro}, \citenamefont {Pricoupenko}, \citenamefont {Astrakharchik},
  \citenamefont {Boronat},\ and\ \citenamefont {Petrov}}]{PhysRevA.97.061605}%
  \BibitemOpen
  \bibfield  {author} {\bibinfo {author} {\bibfnamefont {G.}~\bibnamefont
  {Guijarro}}, \bibinfo {author} {\bibfnamefont {A.}~\bibnamefont
  {Pricoupenko}}, \bibinfo {author} {\bibfnamefont {G.~E.}\ \bibnamefont
  {Astrakharchik}}, \bibinfo {author} {\bibfnamefont {J.}~\bibnamefont
  {Boronat}}, \ and\ \bibinfo {author} {\bibfnamefont {D.~S.}\ \bibnamefont
  {Petrov}},\ }\href {\doibase 10.1103/PhysRevA.97.061605} {\bibfield
  {journal} {\bibinfo  {journal} {Phys. Rev. A}\ }\textbf {\bibinfo {volume}
  {97}},\ \bibinfo {pages} {061605} (\bibinfo {year} {2018})}\BibitemShut
  {NoStop}%
\bibitem [{\citenamefont {Lewenstein}\ \emph {et~al.}(2012)\citenamefont
  {Lewenstein}, \citenamefont {Sanpera},\ and\ \citenamefont
  {Ahufinger}}]{lewenstein2012ultracold}%
  \BibitemOpen
  \bibfield  {author} {\bibinfo {author} {\bibfnamefont {M.}~\bibnamefont
  {Lewenstein}}, \bibinfo {author} {\bibfnamefont {A.}~\bibnamefont {Sanpera}},
  \ and\ \bibinfo {author} {\bibfnamefont {V.}~\bibnamefont {Ahufinger}},\
  }\href@noop {} {\emph {\bibinfo {title} {Ultracold Atoms in Optical Lattices:
  Simulating Quantum Many-body Systems}}}\ (\bibinfo  {publisher} {Oxford
  University Press},\ \bibinfo {address} {Oxford, U.K},\ \bibinfo {year}
  {2012})\BibitemShut {NoStop}%
\bibitem [{\citenamefont {Kuklov}\ \emph
  {et~al.}(2004{\natexlab{a}})\citenamefont {Kuklov}, \citenamefont
  {Prokof'ev},\ and\ \citenamefont {Svistunov}}]{PhysRevLett.92.050402}%
  \BibitemOpen
  \bibfield  {author} {\bibinfo {author} {\bibfnamefont {A.}~\bibnamefont
  {Kuklov}}, \bibinfo {author} {\bibfnamefont {N.}~\bibnamefont {Prokof'ev}}, \
  and\ \bibinfo {author} {\bibfnamefont {B.}~\bibnamefont {Svistunov}},\ }\href
  {\doibase 10.1103/PhysRevLett.92.050402} {\bibfield  {journal} {\bibinfo
  {journal} {Phys. Rev. Lett.}\ }\textbf {\bibinfo {volume} {92}},\ \bibinfo
  {pages} {050402} (\bibinfo {year} {2004}{\natexlab{a}})}\BibitemShut
  {NoStop}%
\bibitem [{\citenamefont {Kuklov}\ \emph
  {et~al.}(2004{\natexlab{b}})\citenamefont {Kuklov}, \citenamefont
  {Prokof'ev},\ and\ \citenamefont {Svistunov}}]{PhysRevLett.92.030403}%
  \BibitemOpen
  \bibfield  {author} {\bibinfo {author} {\bibfnamefont {A.}~\bibnamefont
  {Kuklov}}, \bibinfo {author} {\bibfnamefont {N.}~\bibnamefont {Prokof'ev}}, \
  and\ \bibinfo {author} {\bibfnamefont {B.}~\bibnamefont {Svistunov}},\ }\href
  {\doibase 10.1103/PhysRevLett.92.030403} {\bibfield  {journal} {\bibinfo
  {journal} {Phys. Rev. Lett.}\ }\textbf {\bibinfo {volume} {92}},\ \bibinfo
  {pages} {030403} (\bibinfo {year} {2004}{\natexlab{b}})}\BibitemShut
  {NoStop}%
\bibitem [{\citenamefont {Trefzger}\ \emph {et~al.}(2009)\citenamefont
  {Trefzger}, \citenamefont {Menotti},\ and\ \citenamefont
  {Lewenstein}}]{PhysRevLett.103.035304}%
  \BibitemOpen
  \bibfield  {author} {\bibinfo {author} {\bibfnamefont {C.}~\bibnamefont
  {Trefzger}}, \bibinfo {author} {\bibfnamefont {C.}~\bibnamefont {Menotti}}, \
  and\ \bibinfo {author} {\bibfnamefont {M.}~\bibnamefont {Lewenstein}},\
  }\href {\doibase 10.1103/PhysRevLett.103.035304} {\bibfield  {journal}
  {\bibinfo  {journal} {Phys. Rev. Lett.}\ }\textbf {\bibinfo {volume} {103}},\
  \bibinfo {pages} {035304} (\bibinfo {year} {2009})}\BibitemShut {NoStop}%
\bibitem [{\citenamefont {Valiente}\ \emph {et~al.}(2010)\citenamefont
  {Valiente}, \citenamefont {Petrosyan},\ and\ \citenamefont
  {Saenz}}]{PhysRevA.81.011601}%
  \BibitemOpen
  \bibfield  {author} {\bibinfo {author} {\bibfnamefont {M.}~\bibnamefont
  {Valiente}}, \bibinfo {author} {\bibfnamefont {D.}~\bibnamefont {Petrosyan}},
  \ and\ \bibinfo {author} {\bibfnamefont {A.}~\bibnamefont {Saenz}},\ }\href
  {\doibase 10.1103/PhysRevA.81.011601} {\bibfield  {journal} {\bibinfo
  {journal} {Phys. Rev. A}\ }\textbf {\bibinfo {volume} {81}},\ \bibinfo
  {pages} {011601} (\bibinfo {year} {2010})}\BibitemShut {NoStop}%
\bibitem [{\citenamefont {Valiente}\ and\ \citenamefont
  {Petrosyan}(2009)}]{Valiente_2009}%
  \BibitemOpen
  \bibfield  {author} {\bibinfo {author} {\bibfnamefont {M.}~\bibnamefont
  {Valiente}}\ and\ \bibinfo {author} {\bibfnamefont {D.}~\bibnamefont
  {Petrosyan}},\ }\href {\doibase 10.1088/0953-4075/42/12/121001} {\bibfield
  {journal} {\bibinfo  {journal} {J. Phys. B: At. Mol. Opt. Phys.}\ }\textbf
  {\bibinfo {volume} {42}},\ \bibinfo {pages} {121001} (\bibinfo {year}
  {2009})}\BibitemShut {NoStop}%
\bibitem [{\citenamefont {Valiente}\ and\ \citenamefont
  {Petrosyan}(2008)}]{Valiente_2008}%
  \BibitemOpen
  \bibfield  {author} {\bibinfo {author} {\bibfnamefont {M.}~\bibnamefont
  {Valiente}}\ and\ \bibinfo {author} {\bibfnamefont {D.}~\bibnamefont
  {Petrosyan}},\ }\href {\doibase 10.1088/0953-4075/41/16/161002} {\bibfield
  {journal} {\bibinfo  {journal} {J. Phys. B: At. Mol. Opt. Phys.}\ }\textbf
  {\bibinfo {volume} {41}},\ \bibinfo {pages} {161002} (\bibinfo {year}
  {2008})}\BibitemShut {NoStop}%
\bibitem [{\citenamefont {Bulgac}(2002)}]{PhysRevLett.89.050402}%
  \BibitemOpen
  \bibfield  {author} {\bibinfo {author} {\bibfnamefont {A.}~\bibnamefont
  {Bulgac}},\ }\href {\doibase 10.1103/PhysRevLett.89.050402} {\bibfield
  {journal} {\bibinfo  {journal} {Phys. Rev. Lett.}\ }\textbf {\bibinfo
  {volume} {89}},\ \bibinfo {pages} {050402} (\bibinfo {year}
  {2002})}\BibitemShut {NoStop}%
\bibitem [{\citenamefont {McGuire}(1964)}]{McGuire64}%
  \BibitemOpen
  \bibfield  {author} {\bibinfo {author} {\bibfnamefont {J.~B.}\ \bibnamefont
  {McGuire}},\ }\href {\doibase 10.1063/1.1704156} {\bibfield  {journal}
  {\bibinfo  {journal} {J. Math. Phys.}\ }\textbf {\bibinfo {volume} {5}},\
  \bibinfo {pages} {622} (\bibinfo {year} {1964})}\BibitemShut {NoStop}%
\bibitem [{\citenamefont {Scott}\ \emph {et~al.}(1994)\citenamefont {Scott},
  \citenamefont {Eilbeck},\ and\ \citenamefont {Gilhøj}}]{SCOTT1994194}%
  \BibitemOpen
  \bibfield  {author} {\bibinfo {author} {\bibfnamefont {A.}~\bibnamefont
  {Scott}}, \bibinfo {author} {\bibfnamefont {J.}~\bibnamefont {Eilbeck}}, \
  and\ \bibinfo {author} {\bibfnamefont {H.}~\bibnamefont {Gilhøj}},\ }\href
  {\doibase https://doi.org/10.1016/0167-2789(94)90115-5} {\bibfield  {journal}
  {\bibinfo  {journal} {Physica D: Nonlinear Phenomena}\ }\textbf {\bibinfo
  {volume} {78}},\ \bibinfo {pages} {194 } (\bibinfo {year}
  {1994})}\BibitemShut {NoStop}%
\bibitem [{\citenamefont {Cazalilla}(2003)}]{PhysRevA.67.053606}%
  \BibitemOpen
  \bibfield  {author} {\bibinfo {author} {\bibfnamefont {M.~A.}\ \bibnamefont
  {Cazalilla}},\ }\href {\doibase 10.1103/PhysRevA.67.053606} {\bibfield
  {journal} {\bibinfo  {journal} {Phys. Rev. A}\ }\textbf {\bibinfo {volume}
  {67}},\ \bibinfo {pages} {053606} (\bibinfo {year} {2003})}\BibitemShut
  {NoStop}%
\bibitem [{\citenamefont {Giuliano}\ \emph {et~al.}(2013)\citenamefont
  {Giuliano}, \citenamefont {Rossini}, \citenamefont {Sodano},\ and\
  \citenamefont {Trombettoni}}]{PhysRevB.87.035104}%
  \BibitemOpen
  \bibfield  {author} {\bibinfo {author} {\bibfnamefont {D.}~\bibnamefont
  {Giuliano}}, \bibinfo {author} {\bibfnamefont {D.}~\bibnamefont {Rossini}},
  \bibinfo {author} {\bibfnamefont {P.}~\bibnamefont {Sodano}}, \ and\ \bibinfo
  {author} {\bibfnamefont {A.}~\bibnamefont {Trombettoni}},\ }\href {\doibase
  10.1103/PhysRevB.87.035104} {\bibfield  {journal} {\bibinfo  {journal} {Phys.
  Rev. B}\ }\textbf {\bibinfo {volume} {87}},\ \bibinfo {pages} {035104}
  (\bibinfo {year} {2013})}\BibitemShut {NoStop}%
\bibitem [{\citenamefont {Cazalilla}\ \emph {et~al.}(2011)\citenamefont
  {Cazalilla}, \citenamefont {Citro}, \citenamefont {Giamarchi}, \citenamefont
  {Orignac},\ and\ \citenamefont {Rigol}}]{RevModPhys.83.1405}%
  \BibitemOpen
  \bibfield  {author} {\bibinfo {author} {\bibfnamefont {M.~A.}\ \bibnamefont
  {Cazalilla}}, \bibinfo {author} {\bibfnamefont {R.}~\bibnamefont {Citro}},
  \bibinfo {author} {\bibfnamefont {T.}~\bibnamefont {Giamarchi}}, \bibinfo
  {author} {\bibfnamefont {E.}~\bibnamefont {Orignac}}, \ and\ \bibinfo
  {author} {\bibfnamefont {M.}~\bibnamefont {Rigol}},\ }\href {\doibase
  10.1103/RevModPhys.83.1405} {\bibfield  {journal} {\bibinfo  {journal} {Rev.
  Mod. Phys.}\ }\textbf {\bibinfo {volume} {83}},\ \bibinfo {pages} {1405}
  (\bibinfo {year} {2011})}\BibitemShut {NoStop}%
\bibitem [{\citenamefont {Giamarchi}(2004)}]{Giamarchi:743140}%
  \BibitemOpen
  \bibfield  {author} {\bibinfo {author} {\bibfnamefont {T.}~\bibnamefont
  {Giamarchi}},\ }\href {\doibase 10.1093/acprof:oso/9780198525004.001.0001}
  {\emph {\bibinfo {title} {{Quantum physics in one dimension}}}},\ Internat.
  Ser. Mono. Phys.\ (\bibinfo  {publisher} {Clarendon Press},\ \bibinfo
  {address} {Oxford},\ \bibinfo {year} {2004})\BibitemShut {NoStop}%
\bibitem [{\citenamefont {Lahaye}\ \emph {et~al.}(2009)\citenamefont {Lahaye},
  \citenamefont {Menotti}, \citenamefont {Santos}, \citenamefont {Lewenstein},\
  and\ \citenamefont {Pfau}}]{Lahaye_2009}%
  \BibitemOpen
  \bibfield  {author} {\bibinfo {author} {\bibfnamefont {T.}~\bibnamefont
  {Lahaye}}, \bibinfo {author} {\bibfnamefont {C.}~\bibnamefont {Menotti}},
  \bibinfo {author} {\bibfnamefont {L.}~\bibnamefont {Santos}}, \bibinfo
  {author} {\bibfnamefont {M.}~\bibnamefont {Lewenstein}}, \ and\ \bibinfo
  {author} {\bibfnamefont {T.}~\bibnamefont {Pfau}},\ }\href {\doibase
  10.1088/0034-4885/72/12/126401} {\bibfield  {journal} {\bibinfo  {journal}
  {Reports on Progress in Physics}\ }\textbf {\bibinfo {volume} {72}},\
  \bibinfo {pages} {126401} (\bibinfo {year} {2009})}\BibitemShut {NoStop}%
\bibitem [{\citenamefont {Trefzger}\ \emph {et~al.}(2011)\citenamefont
  {Trefzger}, \citenamefont {Menotti}, \citenamefont {Capogrosso-Sansone},\
  and\ \citenamefont {Lewenstein}}]{Trefzger_2011}%
  \BibitemOpen
  \bibfield  {author} {\bibinfo {author} {\bibfnamefont {C.}~\bibnamefont
  {Trefzger}}, \bibinfo {author} {\bibfnamefont {C.}~\bibnamefont {Menotti}},
  \bibinfo {author} {\bibfnamefont {B.}~\bibnamefont {Capogrosso-Sansone}}, \
  and\ \bibinfo {author} {\bibfnamefont {M.}~\bibnamefont {Lewenstein}},\
  }\href {\doibase 10.1088/0953-4075/44/19/193001} {\bibfield  {journal}
  {\bibinfo  {journal} {Journal of Physics B: Atomic, Molecular and Optical
  Physics}\ }\textbf {\bibinfo {volume} {44}},\ \bibinfo {pages} {193001}
  (\bibinfo {year} {2011})}\BibitemShut {NoStop}%
\bibitem [{\citenamefont {Arg\"uelles}\ and\ \citenamefont
  {Santos}(2007)}]{PhysRevA.75.053613}%
  \BibitemOpen
  \bibfield  {author} {\bibinfo {author} {\bibfnamefont {A.}~\bibnamefont
  {Arg\"uelles}}\ and\ \bibinfo {author} {\bibfnamefont {L.}~\bibnamefont
  {Santos}},\ }\href {\doibase 10.1103/PhysRevA.75.053613} {\bibfield
  {journal} {\bibinfo  {journal} {Phys. Rev. A}\ }\textbf {\bibinfo {volume}
  {75}},\ \bibinfo {pages} {053613} (\bibinfo {year} {2007})}\BibitemShut
  {NoStop}%
\bibitem [{\citenamefont {{Jepsen}}\ \emph {et~al.}(2020)\citenamefont
  {{Jepsen}}, \citenamefont {{Amato-Grill}}, \citenamefont {{Dimitrova}},
  \citenamefont {{Ho}}, \citenamefont {{Demler}},\ and\ \citenamefont
  {{Ketterle}}}]{2020arXiv200509549J}%
  \BibitemOpen
  \bibfield  {author} {\bibinfo {author} {\bibfnamefont {N.}~\bibnamefont
  {{Jepsen}}}, \bibinfo {author} {\bibfnamefont {J.}~\bibnamefont
  {{Amato-Grill}}}, \bibinfo {author} {\bibfnamefont {I.}~\bibnamefont
  {{Dimitrova}}}, \bibinfo {author} {\bibfnamefont {W.~W.}\ \bibnamefont
  {{Ho}}}, \bibinfo {author} {\bibfnamefont {E.}~\bibnamefont {{Demler}}}, \
  and\ \bibinfo {author} {\bibfnamefont {W.}~\bibnamefont {{Ketterle}}},\
  }\href@noop {} {\bibfield  {journal} {\bibinfo  {journal} {arXiv e-prints}\
  ,\ \bibinfo {eid} {2005.09549}} (\bibinfo {year} {2020})},\ \Eprint
  {http://arxiv.org/abs/2005.09549} {arXiv:2005.09549 [cond-mat.quant-gas]}
  \BibitemShut {NoStop}%
\bibitem [{\citenamefont {Hauschild}\ and\ \citenamefont
  {Pollmann}(2018)}]{10.21468/SciPostPhysLectNotes.5}%
  \BibitemOpen
  \bibfield  {author} {\bibinfo {author} {\bibfnamefont {J.}~\bibnamefont
  {Hauschild}}\ and\ \bibinfo {author} {\bibfnamefont {F.}~\bibnamefont
  {Pollmann}},\ }\href {\doibase 10.21468/SciPostPhysLectNotes.5} {\bibfield
  {journal} {\bibinfo  {journal} {SciPost Phys. Lect. Notes}\ ,\ \bibinfo
  {pages} {5}} (\bibinfo {year} {2018})}\BibitemShut {NoStop}%
\end{thebibliography}%

\clearpage
\appendix
\section*{Effective Hamiltonian of dimers}
{\bf Effective dimer hopping.}
First we study the symmetric problem of two bosons $N_a=N_b=1$ by Eq.~\eqref{eq:TwoBH} in an infinite lattice $L\rightarrow \infty$. In the strong interacting regime $U/t\gg 1$ and $r\equiv \frac{U+U_{AB}}{U}\ll 1$ we can work in the effective Hilbert subspace of dimers made of pairs of bosons $a$ and $b$ localized in the same site~\cite{PhysRevLett.92.050402,PhysRevLett.92.030403,PhysRevLett.103.035304}. The number of these states is given by $L$ and we use the notation $|D_n\rangle$ to denote a pair of bosons $ab$ located at site $n$. The matrix elements of the effective Hamiltonian are given by,
\begin{equation}
\begin{split}
&\langle \alpha |H_{\text{eff}} | \beta \rangle = \langle \alpha |H_{0} | \beta \rangle \\&- \frac{1}{2}\sum_{\gamma} \langle \alpha |H_{t} | \gamma \rangle \langle \gamma |H_{t} | \beta \rangle \left( \frac{1}{E_{\gamma}^0 - E_{\alpha}^0} + \frac{1}{E_{\gamma}^0 - E_{\beta}^0} \right),
\label{Eq:elements}
\end{split}
\end{equation}
where $H_0$ contains the interaction part and $H_t$ the hopping one from the original Hamiltonian~\eqref{eq:TwoBH} and $|\gamma\rangle$ are the set of states outside of the effective Hilbert space which are connected with this one by hopping processes. In our situation these excited states consist of breaking the bosonic pair via moving one of the two bosons to an adjacent site and they have an energy $E_{\gamma}^0=0$. 
Given the interaction part of the subspace $\langle D_n|H_0|D_m\rangle=\delta_{n,m}U_{AB}$ we obtain for the matrix elements of the effective Hamiltonian,
\begin{eqnarray}
\langle D_n|H_{\text{eff}}|D_m\rangle &=& \delta_{n,m} \left(U_{AB} + \frac{4t^2}{U_{AB}} \right)\,\nonumber\\&+& \frac{2t^2}{U_{AB}}\delta_{n+1,m}+ \frac{2t^2}{U_{AB}}\delta_{n-1,m}\,.
\end{eqnarray}
These matrix elements can be identified with a single particle hopping between two adjacent sites with an effective hopping $J^{(2)}=2t^2/U_{AB}\approx -2t^2(1+r)/U$ and an effective chemical potential $-\mu=U_{AB}+4t^2/U_{AB}\approx U(r-1) -4t^2(1+r)/U$, where we expand for $r\ll1$. 

\begin{figure}[t!]
    \centering
    \includegraphics[width=1\columnwidth]{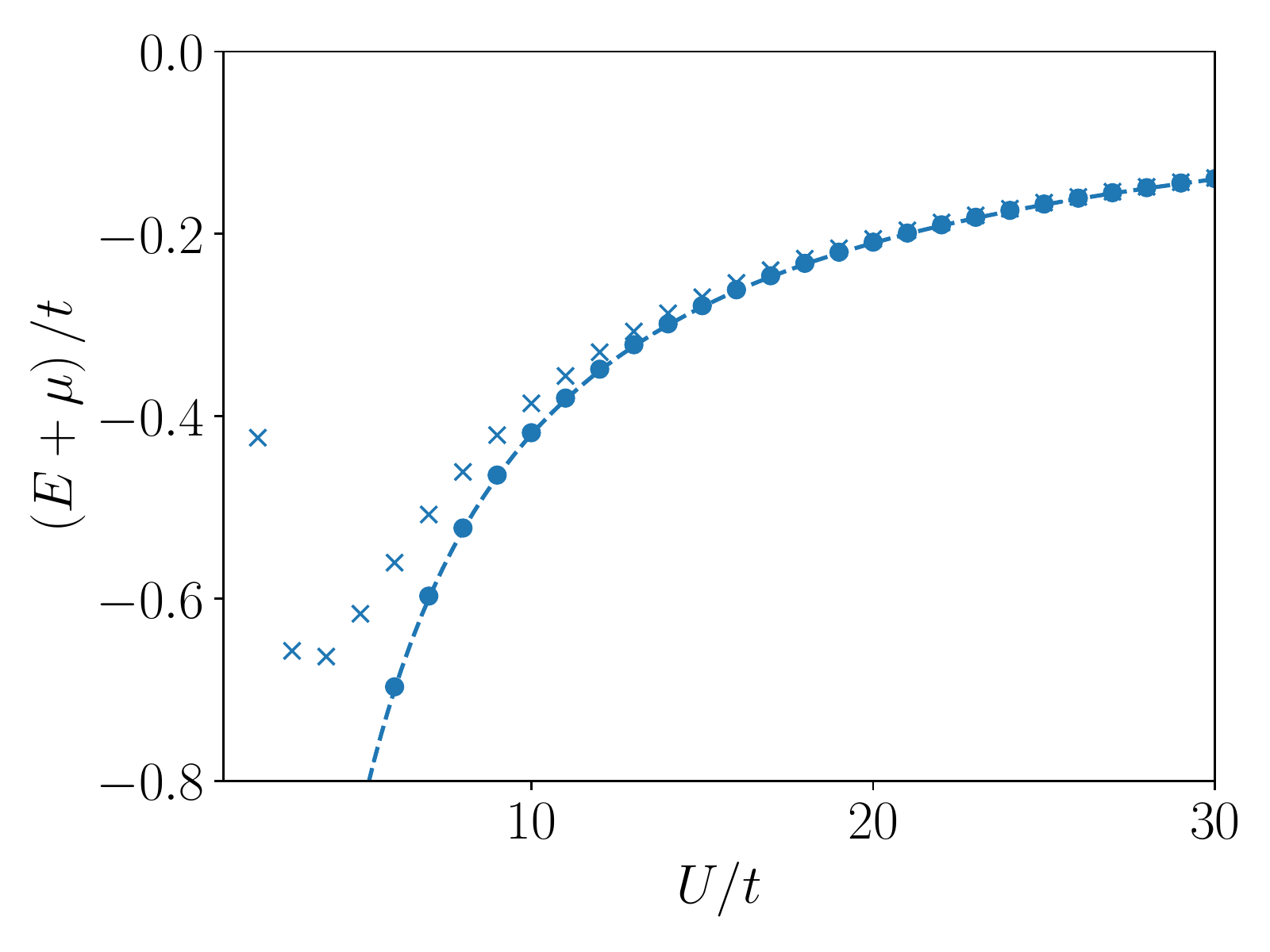}
    \caption{Energy obtained for two bosons $N_A=N_B=1$ using DMRG for the full Hamiltonian (crosses) and the effective one (dots) for $L=32$. Dashed line represents the tight binding result $L\rightarrow \infty$ for a single dimer $E+\mu=2J^{(2)}$.}
    \label{fig:Energy1Dimer}
\end{figure}

{\bf Effective dimer-dimer interaction.}
In order to extract the effective interaction between dimers we study the symmetric problem of four bosons $N_a=N_b=2$. The effective Hilbert subspace is spanned by the set of states $|D_n,D_m\rangle$, with interaction zero order energies $\langle D_n,D_m|H_0|D_n,D_m\rangle=2U_{AB} + 2\delta_{m,n}\left(U_{AB}+U \right)$. The effective interaction is extracted from computing the diagonal matrix elements connected by Eq.~\eqref{Eq:elements}. By computing these matrix elements and performing an expansion for $r\ll 1$ we obtain,
\begin{eqnarray}
\langle  D_n ,D_n|H_{\text{eff}}|D_n,D_n\rangle& =& 2U(2r-1)  - \frac{8t^2}{U}(1+2r)\,  \\ 
\langle  D_{n+1} ,D_n|H_{\text{eff}}|D_{n+1},D_n\rangle  &=& 2U(r-1) -\frac{8t^2}{U} -\frac{4t^2}{U}(1+r)\,\nonumber
\label{Eq:IntElements}
\end{eqnarray}
In order to properly identify the interaction part of the effective Hamiltonian we have to remove the chemical potential contribution from these diagonal elements. The first matrix element in Eq.~\eqref{Eq:IntElements} corresponds to an on-site dimer interaction. The second one corresponds to a nearest-neighbor dimer interaction. By properly identifying the matrix elements we can write the effective Hamiltonian in an operational form
\begin{eqnarray}
H_{\text{eff}}^D+\mu N^D&=&-J^{(2)}\sum_n\left( \hat{c}_{n}^{\dagger}\hat{c}_{n+1}+\text{h.c.}\right) \,\label{Eq:EffHam2}\\
&+&\frac{U^{(2)}}{2}\sum_n \hat{N}_n^D\hat{N}_n^D
+V^{(2)}\sum_n \hat{N}_n^D\hat{N}_{n+1}^D \,,\nonumber
\end{eqnarray}
where $\hat{N}_n^D|N^D_n\rangle=\frac{\hat{n}_{n,a} + \hat{n}_{n,b}}{2}|N^D_n\rangle=N^D_n|N^D_n\rangle$ is the dimer number operator and $\hat{c}^{\dagger}_n$, $\hat{c}_n$ are the respective dimer creation and annihilation operators which satisfy $\left[ \hat{c}_n, \hat{c}^{\dagger}_m\right]=\delta_{n,m}$. The first term describes the hopping of the dimers with an strength $J^{(2)}=2t^2(1+r)/U$. Finally, we have on-site interactions between two dimers with strength $U^{(2)}=Ur-4t^2r/U$ and a nearest-neighbor interaction $V^{(2)}=-4t^2 (1-r)/U$.

\begin{figure}[t!]
    \centering
    \includegraphics[width=1\columnwidth]{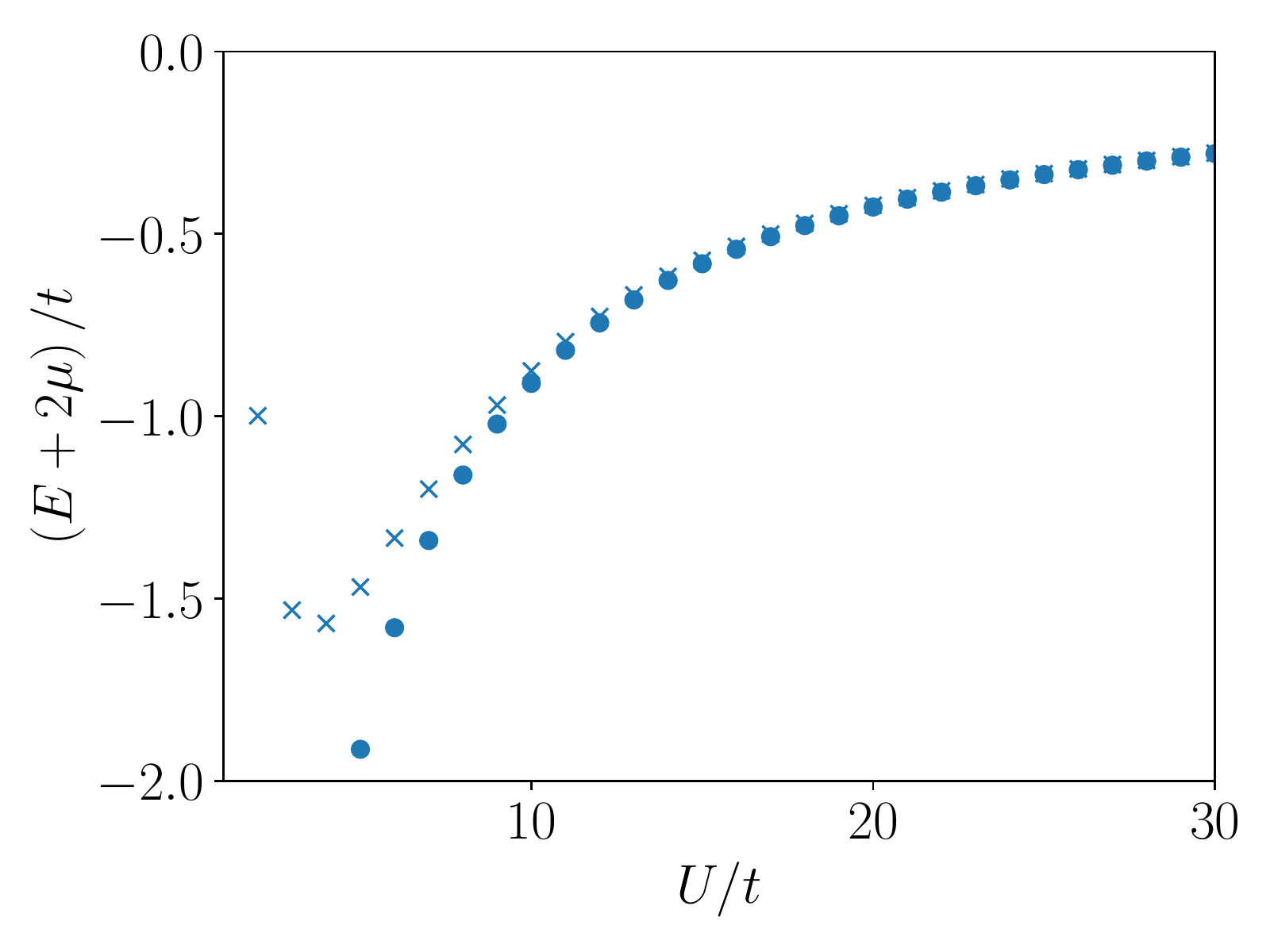}
    \caption{Energy obtained for four bosons $N_A=N_B=2$ using DMRG for the full Hamiltonian (crosses) and the effective one (dots) for $L=32$.}
    \label{fig:Energy1Dimer2}
\end{figure}

\end{document}